\documentclass[aps,prc,preprint,groupedaddress,showpacs,floatfix]{revtex4-1}
\usepackage{graphicx,amsmath,amssymb,color,here}

\newcommand{\nc}{\newcommand}           
\nc{\red}[1]    {\textcolor{red}{#1}}  
\nc{\blue}[1]   {\textcolor{blue}{#1}}  
\nc{\green}[1]   {\textcolor{green}{#1}}  

\begin{document}


\title{Coulomb breakup reactions of $^{11}$Li in the coupled-channel $^9$Li~+~$n$~+~$n$ model}


\author{Yuma Kikuchi}
\email[]{yuma@rcnp.osaka-u.ac.jp}
\affiliation{Research Center for Nuclear Physics (RCNP), Osaka University, Ibaraki 567-0047, Japan}

\author{Takayuki Myo}
\affiliation{General Education, Faculty of Engineering, Osaka Institute of Technology, Osaka 535-8585, Japan}
\affiliation{Research Center for Nuclear Physics (RCNP), Osaka University, Ibaraki 567-0047, Japan}

\author{Kiyoshi Kat\=o}
\affiliation{Nuclear Reaction Data Centre, Faculty of Science, Hokkaido University, Sapporo 060-0810, Japan}

\author{Kiyomi Ikeda}
\affiliation{Nishina Center of Accelerator-Based Sciences, The Institute of Physical and Chemical Research (RIKEN), Wako 351-0198, Japan}


\date{\today}

\begin{abstract}
We investigate the three-body Coulomb breakup of a two-neutron halo nucleus $^{11}$Li.
We use the coupled-channel $^9$Li~+~$n$~+~$n$ three-body model, which includes the coupling between last neutron states and the various $2p$-$2h$ configurations in $^9$Li due to the tensor and pairing correlations.
The three-body scattering states of $^{11}$Li are described by using the combined methods of the complex scaling and the Lippmann-Schwinger equation.
The calculated breakup cross section successfully reproduces the experiments.
The large mixing of the $s$-state in the halo ground state of $^{11}$Li is shown to play an important role in explanation of shape and strength of the breakup cross section.
In addition, we predict the invariant mass spectra for binary subsystems of $^{11}$Li.
It is found that the two kinds of virtual $s$-states of $^9$Li-$n$ and $n$-$n$ systems in the final three-body states of $^{11}$Li largely contribute to make low-lying peaks in the invariant mass spectra.
On the other hand, in the present analysis, it is suggested that the contributions of the $p$-wave resonances of $^{10}$Li is hardly confirmed in the spectra.
\end{abstract}


\maketitle

\section{Introduction}
Recent development of RI beam experiments reveals exotic and interesting properties of unstable nuclei beyond the stability line~\cite{Tanihata96}.
In particular, the neutron halo structure is one of the most interesting topics, and many studies have been performed from both experimental~\cite{Tanihata96} and theoretical~\cite{Zhukov93} sides since the discovery of the halo structure in $^{11}$Li~\cite{Tanihata85}.
The halo nuclei, such as $^6$He, $^{11}$Li, $^{11}$Be, and $^{14}$Be, have been observed to have large matter radii~\cite{Tanihata96} and narrow momentum distributions of the last neutrons~\cite{Kobayashi92}.
These facts reflect the ground-state structure in halo nuclei, in which the weakly-bound last neutrons are spread widely far from the core nucleus.

In addition to the exotic structure in the ground states, breakup reactions of halo nuclei are expected to provide us with much information on the excited states, which are mostly located above the particle thresholds.
The transition mechanism into the excited states is an important issue to be understood. 
Experimentally, the Coulomb breakup reactions have been performed to search for the electric dipole responses of halo nuclei~\cite{Ikeda88}.
The observed cross sections commonly show the low-lying enhancement above the breakup thresholds~\cite{Fukuda04,Aumann99,Wang02,Ieki93,Shimoura95,Zinser97,Nakamura07}, and this enhancement has been considered to be related to the exotic halo structure.
For one-neutron halo cases such as $^{11}$Be~\cite{Fukuda04}, the mechanism of the Coulomb breakup reaction has been discussed as the direct breakup from the halo ground state to the non-resonant continuum states.
The low-lying enhancement can be interpreted as the reflection of the weakly binding halo structure.
On the other hand, for two-neutron halo cases such as $^6$He and $^{11}$Li, the breakup mechanism is complicated because the two-neutron halo nuclei are broken up to the core~+~$n$~+~$n$ three-body scattering states due to the Borromean nature and these final scattering states contain various kinds of correlations, such as of the binary subsystems of core-$n$ and $n$-$n$~\cite{Esbensen92,Kikuchi09,Kikuchi10}.

It has been observed in $^{11}$Li that the amount of the $(1s_{1/2})^2$ component (45$\pm$10 \%) of the halo neutrons is comparable to that of $(0p_{1/2})^2$~\cite{Simon99}.
This fact indicates the breaking of the $N=8$ magic number in the ground state of $^{11}$Li.
For $^{10}$Li, the virtual $s$-state is suggested from a large negative value of the observed scattering length~\cite{Simon07,Jeppesen06}.
The experiments on $^{10}$Li also support the existence of $p$-wave resonances at low excitation energies~\cite{Jeppesen06}.
Based on those observed properties of $^{10}$Li and $^{11}$Li, it is necessary to understand the breakup mechanism of $^{11}$Li using the Coulomb response.

Theoretically, the exotic properties of $^{11}$Li have been studied by using several kinds of approaches, such as the $^9$Li~+~$n$~+~$n$ three-body models~\cite{Zhukov93,Esbensen92,Ikeda92}, no-core shell model~\cite{Navratil09}, fermionic molecular dynamics~\cite{Nortershauser11}, antisymmetrized molecular dynamics~\cite{Enyo95}, stochastic variational method~\cite{Varga02}, microscopic cluster model~\cite{Descouvemont97}, and so on.
For the Coulomb breakup, three-body models have been often used to investigate the mechanism of the breakup into the three-body final states.
Most of the studies based on the three-body models assume to make the $1s_{1/2}$ orbit degenerated with the $0p_{1/2}$ one energetically by using the different $^9$Li-$n$ interactions for even and odd parity states~\cite{Thompson94,Hagino09}.
However, those models cannot reproduce the observed large $s$-wave mixing in $^{11}$Li~\cite{Hagino09}.
Furthermore, Esbensen {\it et al.}~\cite{Esbensen07} have mentioned that the $^9$Li~+~$n$~+~$n$ three-body model fails in explanation of the observed charge radius and the dipole strength in $^{11}$Li consistently.
In order to overcome these difficulties in $^{11}$Li, a new approach beyond the $^9$Li~+~$n$~+~$n$ three-body model has been desired.

Myo {\it et al.}~\cite{Myo07,Myo08,Ikeda10} have attempted to understand the exotic properties of $^{11}$Li by taking into account the tensor and pairing correlations in $^9$Li using the tensor-optimized shell model (TOSM)~\cite{Myo09,Myo11}.
In their approach, the $s$-wave virtual states and the $p$-wave resonances in $^{10}$Li are reproduced simultaneously by taking into account the $2p$-$2h$ excitations of the $^9$Li core coming from the tensor and pairing correlations.
The Pauli principle between the last neutron and the $^9$Li core plays an important role. 
This dynamical description has also been applied to the $^{11}$Li system by using the coupled-channel $^9$Li~+~$n$~+~$n$ model.
In this model, the tensor correlation produces the specific $2p$-$2h$ excitation from the $0s_{1/2}$ orbit to the $0p_{1/2}$ orbit~\cite{Myo09}, and the Pauli-blocking between the excited $0p_{1/2}$ neutron in the $^9$Li core and the last halo neutrons dynamically pushes up the energy of the $0p_{1/2}$ orbit in $^{11}$Li~\cite{Myo07,Myo08}.
As a result, the energies of the $p$-wave and $s$-wave states of $^{11}$Li get close to each other, and a large $s$-wave mixing in the $^{11}$Li ground state is brought.
Furthermore, this coupled-channel $^9$Li~+~$n$~+~$n$ model is shown to reproduce various physical quantities such as the matter and charge radii, and the dipole strength of $^{11}$Li consistently. 

In this study, we investigate the Coulomb breakup reaction of $^{11}$Li by using the coupled-channel $^9$Li~+~$n$~+~$n$ three-body model proposed in Refs.~\cite{Myo07,Myo08}.
The aim of this work is to investigate the role of the dynamical mixing of $(1s_{1/2})^2$-configuration in the ground state on the reaction mechanism of the three-body Coulomb breakup of $^{11}$Li.
The other aim is to get the knowledge of the binary correlations of $^9$Li-$n$ and $n$-$n$ in the breakup final states.
For these purpose, it is necessary to describe the three-body scattering states with a correct boundary condition.
This problem is solved by applying the complex-scaled solutions of the Lippmann-Schwinger equation (CSLS)~\cite{Kikuchi09,Kikuchi10}. 
This method has been recently developed and successfully applied to the Coulomb breakup reactions of $^6$He and the $\alpha$~+~$d$ scattering using the $\alpha$~+~$p$~+~$n$ model for $^6$Li ~\cite{Kikuchi11}. 

The Coulomb breakup cross sections have been analyzed as functions of subsystem energies to understand the correlations in two-neutron halo nuclei~\cite{Esbensen92,Chulkov05,Ershov06,Hagino09,Kikuchi09,Kikuchi10}.
For two-neutron halo nucleus $^6$He, the breakup mechanism of $^6$He has been analyzed based on the $\alpha$~+~$n$~+~$n$ three-body model, in which the employed $\alpha$-$n$ interaction reproduces the observed phase shifts but is not enough strong to produce the virtual $s$-states.
In our previous works of $^6$He ~\cite{Kikuchi09,Kikuchi10}, it was found that the characteristic shape of the breakup cross section is weakly influenced by the halo structure in the ground state.
The magnitude and the peak position of the cross section can be explained with a dominant contribution from the final state interactions (FSI) which produce the $^5$He(3/2$^-$) resonance and the $n$-$n$ virtual state. 

In contrast to $^6$He, the Coulomb breakup cross sections of $^{11}$Li have been shown to depend strongly on the structure, in particular, on the $s$-wave mixing in the halo ground state~\cite{Esbensen92,Hagino09,Myo03}.
The simple $^9$Li~+~$n$~+~$n$ model calculations~\cite{Esbensen92,Hagino09}, which predict a small $s$-wave mixing (about 20 \%), reproduce the low-energy enhancement in the cross section.
However, the energy position of the enhancement is higher than the observed one~\cite{Nakamura07}.
In this paper, it is shown that this serious problem in the $^9$Li~+~$n$~+~$n$ model is solved satisfactory.
Furthermore, the previous $^9$Li~+~$n$~+~$n$ model suggests that the virtual $s$-state and the $p_{1/2}$ resonances of $^{10}$Li are observed in the strength distribution together with the $n$-$n$ correlation as functions of subsystem energies~\cite{Hagino09}.
These problems are also discussed in the present coupled-channel $^9$Li~+~$n$~+~$n$ analysis. 

This paper is organized as follows.
In Sec.~\ref{sec:mod}, we expain the coupled-channel  $^9$Li~+~$n$~+~$n$ three-body model and the formalism of the Coulomb breakup reaction using CSLS.
In Sec.~\ref{sec:res}, we show the results of the Coulomb breakup cross section of $^{11}$Li and the invariant mass spectra of its binary subsystems, and discuss the breakup mechanism of $^{11}$Li.
All results and discussion are summarized in Sec.~\ref{sec:sum}.

\section{Framework\label{sec:mod}}
\subsection{Coupled-channel three-body model for $^{11}$Li}
We give a brief explanation of the coupled-channel $^9$Li~+~$n$~+~$n$ three-body model of $^{11}$Li employed here, the detail of which is given in Refs.~\cite{Myo07,Myo08}.
To solve many-body correlations not only of the last two neutrons but also of the coupling with degrees of freedom in the $^9$Li core, we start with the Schr\"odinger equation:
\begin{equation}
\hat{H}\Phi_{J^\pi} = E\Phi_{J^\pi},
\label{eq:sch}
\end{equation}
where $\hat{H}$ and $\Phi_{J^\pi}$ are the total Hamiltonian and the wave function with spin-parity $J^\pi$ for $^{11}$Li, respectively.

The wave function $\Phi_{J^\pi}$ is described as follows:
\begin{equation}
\Phi_{J^\pi} = \sum_c a_c \mathcal{A} \left[\Phi^c_{3/2^-}(^9\text{Li}) \otimes \chi^c_{J_0} (nn) \right]_{J^\pi},
\end{equation}
where $\mathcal{A}$ is an antisymmetrizer, and $\Phi^c(^9\text{Li})$ and $\chi^c(nn)$ are the wave functions for the $^9$Li core and the last two neutrons, respectively.
The index $c$ represents the quantum numbers of the $^9$Li core configurations.
The channel amplitude $a_c$ is determined by solving the coupling between $^9$Li and the relative wave functions of the last two neutrons using the coupled-channel equation Eq.~(\ref{eq:sch}). 
Using TOSM~\cite{Myo07,Myo08} for the configurations of the $^9$Li core, we employ two kinds of $2p$-$2h$ configurations: $(nn)_{J^{\pi}=0^+}$ for the pairing correlation and $(pn)_{J^{\pi}=1^+}$ for the tensor correlation, in addition to the $0p$-$0h$ one of $(0s)^4(0p_{3/2})^4_{\nu}(0p_{3/2})_{\pi}^1$.
Thus our model is considered to be an extension of the usual $^9$Li~+~$n$~+~$n$ model, because this coupled-channel model is equivalent to the simple $^9$Li~+~$n$~+~$n$ model  when the only $0p$-$0h$ configuration is taken for $^9$Li.
The Pauli principle between the last two neutrons is taken into account by antisymmetrizing the relative wave function $\chi^c(nn)$.
For the $^9$Li-$n$ part, we take into account the Pauli principle by projecting out the Pauli forbidden states occupied by the $^9$Li core from the relative wave function $\chi^c(nn)$.

Here, the total Hamiltonian for the coupled-channel $^9$Li~+~$n$~+~$n$ three-body system is given as
\begin{equation}
\hat{H} = \sum_{i=1}^3 \hat{T}_i - \hat{T}_\text{cm} + V_{n\text{-}n} + \sum_{i=1}^2 \hat{V}_\text{core-$n$} (\mathbf{r}_i) + \hat{h}(^9\text{Li}),
\label{eq:ham}
\end{equation}
where $\hat{T}_i$ and $\hat{T}_\text{cm}$ are kinetic energies of each cluster and the center-of-mass motion of the $^9$Li~+~$n$~+~$n$ three-body system, respectively.
For the $n$-$n$ interaction $\hat{V}_{n\text{-}n}$, we employ the realistic Argonne $v8'$ force~\cite{Wiringa95}.
The relative coordinate $\mathbf{r}_i$ represents that between the $^9$Li core and the $i$-th last neutron, and $\hat{V}_\text{core-$n$}$ is the $^9$Li-$n$ interaction.
For $\hat{V}_\text{core-$n$}$, we employ a potential folding an effective $NN$ interaction with $^9$Li core density~\cite{Kato99}.
We here use the modified Hasegawa-Nagata (MHN) potential~\cite{Hasegawa71,Furutani80} obtained in the $G$-matrix calculation as the effective $NN$ interaction.
We introduce one parameter $\delta$, which enhances the strength of the intermediate range of MHN potential from the original one in order to describe the starting energy dependence dominantly coming from the tensor force in the $G$-matrix calculation~\cite{Myo07}.
In the present calculation, we take this parameter $\delta$ as $0.1893$ to reproduce the observed two-neutron separation energy of $^{11}$Li.

The microscopic internal Hamiltonian for the $^{9}$Li core, $\hat{h}(^9\text{Li})$, is given as
\begin{equation}
\hat{h}(^9\text{Li}) = \sum_{i=1}^9 \hat{t}_i - \hat{t}_\text{cm} + \sum_{i<j}^9 \hat{v}_{ij}.
\label{eq:inham}
\end{equation}
Here, $\hat{t}_i$ and $\hat{t}_\text{cm}$ are the kinetic energies for each nucleon and the center-of-mass motion in the $^9$Li core, respectively.
The two-body $NN$ interaction $\hat{v}_{ij}$, whose detail is given in Ref.~\cite{Myo09}, consists of central, spin-orbit, tensor, and Coulomb terms.
Using the Hamiltonian in Eq.~(\ref{eq:inham}), the $2p$-$2h$ configurations in the $^9$Li core are involved by the pairing and tensor correlations.

\begin{figure}
\centering{\includegraphics[width=7cm,clip]{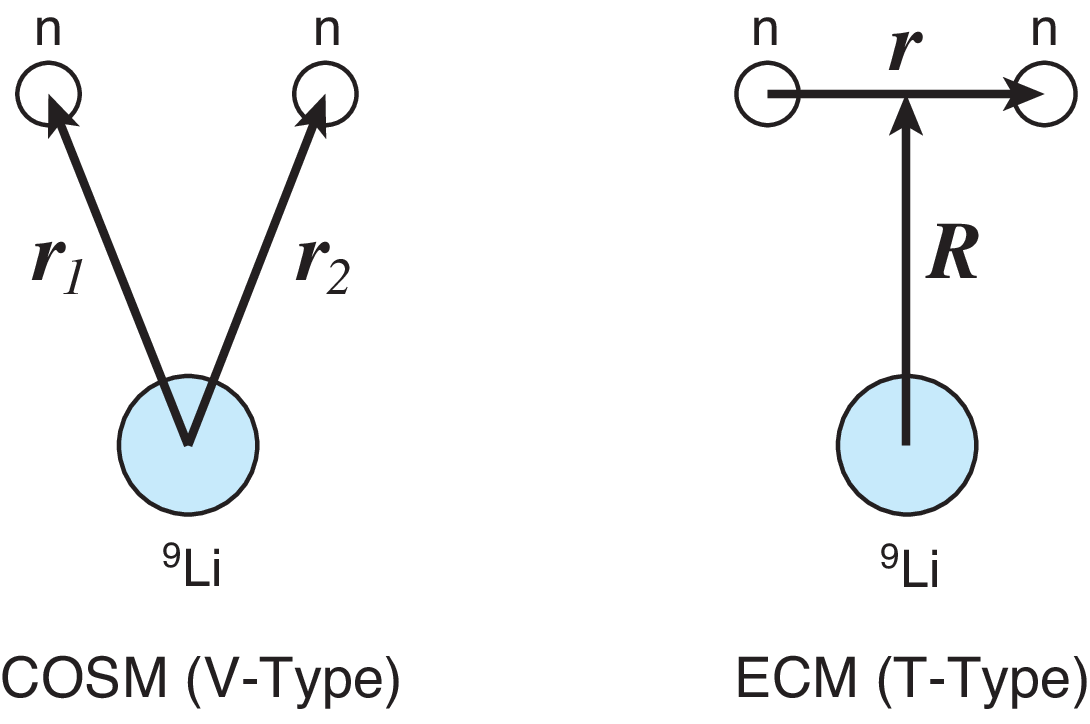}}
\caption{\label{fig:hybVT} (Color online) Coordinate sets in hybrid-VT model for $^{11}$Li.}
\end{figure}
The  wave function $\chi^c(nn)$ of the last two neutrons are solved by using a few-body technique.
Here we employ the variational approach called the hybrid-$VT$ model~\cite{Aoyama95}.
In this model, we expand the relative wave function $\chi^c(nn)$ with two kinds of basis functions, shown in FIG.~\ref{fig:hybVT}: One is the cluster-orbital shell model (COSM; V-type) and the other is the extended cluster model (ECM; T-type).
These two kinds of basis functions are important for expressing the $^9$Li-$n$ and $n$-$n$ correlations simultaneously.
Radial components of each relative wave function are expanded using the Gaussian basis functions~\cite{hiyama03}.
The number of the basis functions is determined to reach the convergence of the numerical results.
In the present calculation, we employ 18 basis functions for each coordinates, and the range of the Gaussian is taken up to 60 fm.

\subsection{Complex-scaled solutions of the Lippmann-Schwinger equation}
We describe the $^9$Li~+~$n$~+~$n$ three-body scattering states of $^{11}$Li by using CSLS~\cite{Kikuchi09} to calculate the Coulomb breakup cross section.
Before going into the formalism of CSLS, we briefly explain CSM~\cite{Aoyama06}.
In CSM, the relative coordinates shown in FIG.~\ref{fig:hybVT} are commonly transformed as follows:
\begin{equation}
U(\theta) \boldsymbol{\xi} U^{-1}(\theta) = \boldsymbol{\xi} e^{i\theta},
\end{equation}
where $U(\theta)$ is a complex scaling operator with a scaling angle $\theta$ being a real number.
The coordinate $\boldsymbol{\xi}$ represents the set of the two relative coordinates in the three-body system as used in FIG.~\ref{fig:hybVT}.
Applying this transformation to the Hamiltonian $\hat{H}$, we obtain the  complex-scaled Schr\"odinger equation is given as
\begin{equation}
\hat{H}^\theta \Phi^\theta = E^\theta \Phi^\theta,
\label{eq:cssch}
\end{equation}
where $\hat{H}^\theta$ and $\Phi^\theta$ are the complex-scaled Hamiltonian and the complex-scaled wave function given as
\begin{equation}
\hat{H}^\theta = U(\theta) \hat{H} U^{-1}(\theta)
\end{equation}
and
\begin{equation}
\Phi^\theta = U(\theta) \Phi(\boldsymbol{\xi}) = e^{\left(\frac{3}{2}i\theta\right)\cdot f}\Phi(\boldsymbol{\xi} e^{i\theta}),
\end{equation}
respectively.
The factor, $e^{\left(\frac{3}{2}i\theta\right)\cdot f}$, comes from the Jacobian in the volume integral, and $f=2$ for the three-body systems.
By solving the complex-scaled Schr\"odinger equation given in Eq.~(\ref{eq:cssch}) with a finite number of $L^2$ basis functions such as Gaussian, we obtain the eigenstates and energy eigenvalues of $\hat{H}^\theta$ as $\{\Phi^\theta_n\}$ and $\{E^\theta_n\}$ with a state index $n$, respectively.

\begin{figure}
\centering{\includegraphics[width=8cm,clip]{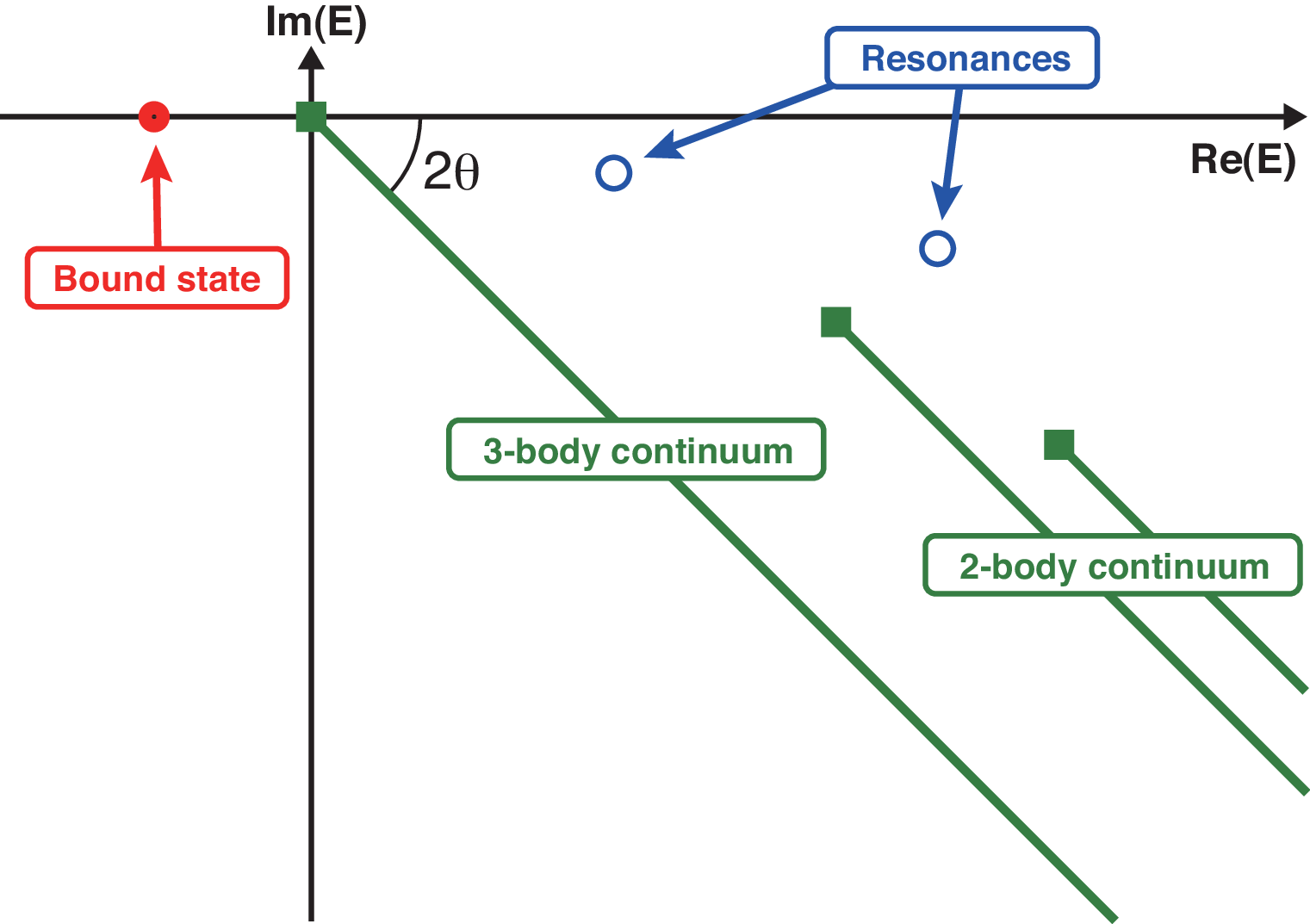}}
\caption{\label{fig:csme} (Color online) Schematic picture of energy eigenvalues of the three-body Borromean system in CSM.}
\end{figure}
All the energy eigenvalues $\{E^\theta_n\}$ are obtained on a complex energy plane, governed by the ABC theorem~\cite{Aguilar71,Balslev71}, and their imaginary parts represent the outgoing boundary conditions.
In FIG.~\ref{fig:csme}, we show a schematic distribution of energy eigenvalues of the three-body Borromean system in CSM.
In CSM, the resonances of a many-body system are obtained as the isolated poles with the $L^2$ basis functions.
On the other hand, the energy eigenvalues of continuum states are obtained on the $2\theta$-rotated branch cuts starting from different thresholds of two- and three-body decay channels, such as $^{10}$Li~+~$n$ and $^9$Li~+~$n$~+~$n$ in the case of $^{11}$Li.
This classification of continuum states in CSM imposes that the outgoing boundary condition for each open channel is taken into account automatically by the imaginary parts of energy eigenvalues.
Using the classification of continuum states in CSM, we can describe three-body scattering states without any explicit enforcement of boundary conditions.

Furthermore, the complex-scaled eigenstates satisfy the extended completeness relation (ECR)~\cite{Myo01,Myo98}, consisting of bound states, resonances, and rotated continua for the state index $n$ as
\begin{equation}
\mathbf{1} = \sum_n\hspace{-0.46cm}\int | \Phi^\theta_n \rangle \langle \tilde{\Phi}^\theta_n |,
\label{eq:ecr}
\end{equation}
where $\{\tilde{\Phi}^\theta_n, \Phi^\theta_n\}$ form a set of biorthogonal states~\cite{Myo01}.
This relation is used when we describe the scattering states with the Lippmann-Schwinger equation.

In CSLS, we start with the formal solution of the Lippmann-Schwinger equation given as
\begin{equation}
\Psi^{(\pm)}(\mathbf{k},\mathbf{K}) = \phi_0(\mathbf{k},\mathbf{K}) + \lim_{\varepsilon \to 0} \frac{1}{E-\hat{H}\pm i\varepsilon}\hat{V} \phi_0(\mathbf{k},\mathbf{K}),
\label{eq:lseq}
\end{equation}
where $\phi_0(\mathbf{k},\mathbf{K})$ is the solution of the asymptotic Hamiltonian $\hat{H}_0$ with two relative momenta $\mathbf{k}$ and $\mathbf{K}$ in Jacobi coordinates of the three-body system.
The total Hamiltonian $\hat{H}$ is the same as given in Eq.~(\ref{eq:ham}), and the interaction $\hat{V}$ is defined by subtracting the asymptotic Hamiltonian $\hat{H}_0$ from $\hat{H}$.

For the three-body breakup reaction of $^{11}$Li, the asymptotic Hamiltonian $\hat{H}_0$ consists of the kinetic part of the $^9$Li~+~$n$~+~$n$ system and the internal Hamiltonian of $^9$Li since no binary subsystems have bound states because of the Borromean condition for $^{11}$Li. 
Hence, we can define $\hat{H}_0$ and its solution $\phi_0$ as
\begin{align}
\hat{H}_0 &= \hat{h}(^9\text{Li})+\sum_{i=1}^3 \hat{t}_i - \hat{T}_\text{cm}, \\
\hat{H}_0 \phi_0 &= \left(\varepsilon_{^9\text{Li}} + \frac{\hbar^2 k^2}{2\mu} + \frac{\hbar^2 K^2}{2M}\right)\phi_0, \\
\langle \mathbf{r},\mathbf{R} | \phi_0(\mathbf{k},\mathbf{K}) \rangle &= \frac{1}{(2\pi)^3} \Phi_\text{gs}(^9\text{Li}) \otimes e^{i\mathbf{k}\cdot\mathbf{r}+i\mathbf{K}\cdot\mathbf{R}},
\end{align}
where $\mu$ and $M$ are the reduced masses corresponding to $\mathbf{k}$ and $\mathbf{K}$, respectively.
The coordinates $\mathbf{r}$ and $\mathbf{R}$ are conjugate to the momenta $\mathbf{k}$ and $\mathbf{K}$, respectively.
Here, we assume that the $^9$Li core is in the ground state with energy of $\varepsilon_{^9\text{Li}}$ after the breakup.
The interaction $\hat{V}$ in Eq.~(\ref{eq:lseq}) is given as
\begin{equation}
\hat{V} = \sum_{i=1}^2 \hat{V}_\text{core-$n$}(\mathbf{r}_i) + \hat{V}_{n\text{-}n}.
\label{eq:int}
\end{equation}

We consider the incoming scattering states in bra-representation, which are used as the final states in the Coulomb breakup cross section of $^{11}$Li.
Assuming the Hermiticities of $\hat{H}$ and $\hat{V}$, Eq.~(\ref{eq:lseq}) is rewritten as
\begin{equation}
\langle \Psi^{(-)}(\mathbf{k},\mathbf{K}) | = \langle \phi_0(\mathbf{k},\mathbf{K}) | + \langle \phi_0(\mathbf{k},\mathbf{K}) | \hat{V} \lim_{\varepsilon\to0} \frac{1}{E-\hat{H}+i\varepsilon}.
\label{eq:insca}
\end{equation}
In CSLS, we express the Green's function in Eq.~(\ref{eq:insca}) in terms of the complex-scaled Green's function.
The complex-scaled Green's function with the outgoing boundary condition, $\mathcal{G}^\theta(E)$, is connected to the non-scaled Green's function $\mathcal{G}(E)$ as follows:
\begin{equation}
\lim_{\varepsilon\to0} \frac{1}{E-\hat{H}+i\varepsilon} = \mathcal{G}(E) = U^{-1}(\theta) \mathcal{G}^\theta(E) U(\theta).
\label{eq:relgf}
\end{equation}
The explicit form of $\mathcal{G}^\theta(E)$ is defined as
\begin{equation}
\mathcal{G}^\theta(E) = \frac{1}{E-\hat{H}^\theta} = \sum_n\hspace{-0.46cm}\int \frac{|\Phi^\theta_n\rangle\langle\tilde{\Phi}^\theta_n|}{E-E^\theta_n},
\label{eq:csgf}
\end{equation}
where ECR defined in Eq.~(\ref{eq:ecr}) is inserted.
Using Eqs.~(\ref{eq:relgf}) and (\ref{eq:csgf}), we obtain the incoming scattering state $\Psi^{(-)}$ in CSLS as
\begin{equation}
\begin{split}
\langle \Psi^{(-)}&(\mathbf{k},\mathbf{K}) |
= \langle \phi_0(\mathbf{k},\mathbf{K}) | \\
&+ \sum_n \hspace{-0.46cm} \int \langle \phi_0(\mathbf{k},\mathbf{K}) | \hat{V} U^{-1}(\theta) | \Phi^\theta_n \rangle \frac{1}{E-E^\theta_n} \langle \tilde{\Phi}^\theta_n | U(\theta).
\end{split}
\label{eq:csls}
\end{equation}

It is noted that the scattering states in Eq.~(\ref{eq:csls}) consist of two terms:
The first term describes the non-interacting three-body continuum state, which is the same as given in Eq.~(\ref{eq:lseq}).
The second term contains all information of FSI.
Using Eq.~(\ref{eq:csls}), we can extract the effect of each component from the breakup cross section.

\subsection{Coulomb breakup cross section}
The Coulomb breakup reaction is considered to be dominated by the $E1$ transition.
In the present calculation, we obtain the Coulomb breakup cross section using the $E1$ transition strength and the virtual photon number from the equivalent photon method~\cite{Hoffmann84,Bertulani88}.

In CSLS, we calculate the momentum distribution of the $E1$ transition strength from the $^{11}$Li ground state into the $^9$Li~+~$n$~+~$n$ three-body scattering states.
The distribution is given as
\begin{equation}
\frac{d^6B(E1)}{d\mathbf{k}d\mathbf{K}} = \frac{1}{2J_\text{gs}+1}
\left|\langle \Psi^{(-)}(\mathbf{k},\mathbf{K}) || \hat{O}(E1) || \Phi_\text{gs} \rangle \right|^2,
\label{eq:mom_dis}
\end{equation}
where $\Phi_\text{gs}$ and $J_\text{gs}$ are the wave function and the total spin for the initial ground state of $^{11}$Li, respectively.
The wave function $\Psi^{(-)}(\mathbf{k},\mathbf{K})$ is that for the three-body scattering state of $^9$Li~+~$n$~+~$n$ with relative momenta, $\mathbf{k}$ and $\mathbf{K}$, and here, is described by using CSLS as shown in Eq.~(\ref{eq:csls}).
Using the coordinate sets shown in FIG.~\ref{fig:hybVT}, the $E1$ operator $\hat{O}(E1)$ is given as
\begin{equation}
\hat{O}_m(E1) = \frac{6}{11}eRY_{1m}(\hat{\mathbf{R}}) = \frac{3}{11}\left(r_1Y_{1m}(\hat{\mathbf{r}}_1)+r_2Y_{1m}(\hat{\mathbf{r}}_2)\right),
\end{equation}
where $Y_{1m}$ is the spherical harmonics and $m$ is the $z$ component of the operator. 

Using Eq.~(\ref{eq:mom_dis}), the two-dimensional energy distribution is defined as
\begin{equation}
\begin{split}
\frac{d^2B(E1)}{d\varepsilon_1d\varepsilon_2} = &\iint d\mathbf{k} d\mathbf{K}
\frac{d^6B(E1)}{d\mathbf{k}d\mathbf{K}} \\
&\times \delta\left(\varepsilon_1 - \frac{\hbar^2k^2}{2\mu}\right) \delta\left(\varepsilon_2 - \frac{\hbar^2K^2}{2M}\right),
\end{split}
\label{eq:2d_ene}
\end{equation}
where $\varepsilon_1$ and $\varepsilon_2$ are the relative energies of each subsystem, such as $^9$Li-$n$ and $n$-$n$.
Similarly, the total energy distribution of the $E1$ strength is given as
\begin{equation}
\begin{split}
\frac{dB(E1)}{dE} = &\iint d\mathbf{k} d\mathbf{K} \frac{d^6B(E1)}{d\mathbf{k} d\mathbf{K}}\\
&\times \delta\left(E-\varepsilon_{^9\text{Li}}-\frac{\hbar^2k^2}{2\mu}-\frac{\hbar^2K^2}{2M}\right).
\end{split}
\label{eq:tot_ene}
\end{equation}

Using Eqs.~(\ref{eq:2d_ene}) and (\ref{eq:tot_ene}) and the equivalent photon method, we can calculate the Coulomb breakup cross sections.
The two-dimensional energy distribution and total energy distribution of the cross sections are given as
\begin{equation}
\frac{d^2\sigma}{d\varepsilon_1 d\varepsilon_2} = \frac{16\pi^3}{9\hbar c} N_{E1}(E_\gamma) \frac{d^2B(E1)}{d\varepsilon_1 d\varepsilon_2}
\label{eq:2d_cs}
\end{equation}
and
\begin{equation}
\frac{d\sigma}{dE} = \frac{16\pi^3}{9\hbar c} N_{E1}(E_\gamma) \frac{dB(E1)}{dE},
\label{eq:tot_cs}
\end{equation}
respectively.
Here, $N_{E1}(E_\gamma)$ is the virtual photon number with the photon energy $E_\gamma$, and $E_\gamma$ is given as
\begin{equation}
E_\gamma = \varepsilon_1 + \varepsilon_2 + S_{2n} = E - \varepsilon_{^9\text{Li}}+ S_{2n},
\end{equation}
where $S_{2n}$ is the two-neutron separation energy of $^{11}$Li.

From Eq.~(\ref{eq:2d_cs}), the invariant mass spectra for binary subsystems such as of $^9$Li-$n$ and $n$-$n$ are given as
\begin{equation}
\frac{d\sigma}{d\varepsilon_1} = \int d\varepsilon_2 \frac{d^2\sigma}{d\varepsilon_1 d\varepsilon_2},
\label{eq:inv}
\end{equation}
where $\varepsilon_1$ is the relative energy of the binary subsystem.

\section{Results\label{sec:res}}
\subsection{Properties of $^{10}$Li and $^{11}$Li}
\begin{table}[t]
\caption{\label{tab:10Li}
Resonance energies and decay widths of $p_{1/2}$ resonances of $^{10}$Li, measured from the $^9$Li~+~$n$ threshold.
The scattering lengths $a_s$ are for the $s_{1/2}$ virtual states of $^{10}$Li.
The experimental data for $a_s$ is also listed.}
\begin{ruledtabular}
\begin{tabular}{cccccc}
& $1^+$ & $2^+$ & $1^-$ & $2^-$ & Exp.\\
\hline
Energy (keV) & 275 & 506 & & & \\
Width (keV) & 150 & 388 & & & \\
$a_s$ (fm) & & & $-6.8$ & $-45.0$ & $-30^{+12}_{-31}$\footnote{Reference~\cite{Simon07}}\\
\end{tabular}
\end{ruledtabular}
\end{table}
\begin{table}[t]
\caption{\label{tab:gspro}Ground-state properties of $^{11}$Li. See the text for details.}
\begin{ruledtabular}
\begin{tabular}{ccc}
& Theor. & Exp. \\
\hline
$S_{2n}$ (keV) & 377 & 378$\pm$5\footnote{Reference~\cite{Bachelet08}} \\
$R_\text{m}$ (fm) & 3.39 & 3.12$\pm$0.16\footnote{Reference~\cite{Tanihata88}} \\
& & 3.53$\pm$0.06\footnote{Reference~\cite{Tostevin97}} \\
& & 3.71$\pm$0.20\footnote{Reference~\cite{Dobrovolsky06}} \\
$R_\text{ch}$ (fm) & 2.43 & 2.467$\pm$0.037\footnote{Reference~\cite{Sanchez06}}\\
& & 2.423$\pm$0.034\footnote{Reference~\cite{Puchalski06}} \\
\hline
$P((s_{1/2})^2) (\%)$  & 44.0 & 45$\pm$10\footnote{Reference~\cite{Simon99}} \\
$P((p_{3/2})^2) (\%)$ & 2.5 & -- \\
$P((p_{1/2})^2) (\%)$ & 46.9 & -- \\
$P((d_{5/2})^2) (\%)$ & 3.1 & -- \\
$P((d_{3/2})^2) (\%)$ & 1.7 & -- \\
\end{tabular}
\end{ruledtabular}
\end{table}

We here discuss the properties of $^{10}$Li and $^{11}$Li obtained in the present coupled-channel $^9$Li~+~$n$~+~$n$ model.
In TABEL~\ref{tab:10Li}, we show the $^{10}$Li results.
We obtain the $p$-wave doublet resonances, $1^+$ and $2^+$, constructed from the coupling between the spins of the last neutron and $^9$Li($3/2^-$).
Two resonances are located at 275 keV and 506 keV measured from the $^9$Li~+~$n$ threshold with the decay widths of 150 keV and 388 keV for $1^+$ and $2^+$, respectively.
There are also the $s$-wave doublet states of $1^-$ and $2^-$ in $^{10}$Li in the same reason of the spin coupling as the $p$-wave case.
The scattering length $a_s$ for the $2^-$ state is obtained as $-45.0$ fm, quite a large negative value, and shows a good agreement with the experimental one within the error bounds~\cite{Simon07}.
The present large negative value of $a_s$ indicates the existence of the virtual $s$-states in $^{10}$Li($2^-$).
It is noted that the previous value of $a_s$ is about $-18$ fm in Ref.~\cite{Myo08}, which is negatively smaller than the present one, because the smaller $2n$ separation energy of $^{11}$Li, suggested from the old experiment, was adopted to determine the $^9$Li-$n$ interaction.

We show the ground-state properties of $^{11}$Li in TABLE~\ref{tab:gspro}.
The calculated matter and charge radii, $R_\text{m}$ and $R_\text{ch}$, are obtained as 3.39 fm and 2.43 fm, respectively, and are in a good agreement with the observed data.
We also calculate the probabilities of the partial wave components $P((lj)^2)$ of the last two neutrons.
The probability of $(s_{1/2})^2$ is very large and consistent with the observed data~\cite{Simon99}, which is almost equal to that of the $(p_{1/2})^2$ component.
The large $s$-wave mixing generates the neutron halo structure and also indicates the breaking of the $N=8$ magic number. 
From the results, the present coupled-channel $^9$Li~+~$n$~+~$n$ model using TOSM for the $^9$Li core well reproduces the structures of $^{10}$Li and $^{11}$Li, in particular, those related to the $s$-wave properties.
Other ground-state properties, such as the quadrupole moment and the spin dipole moment of $^9$Li and $^{11}$Li, have been discussed, and it was shown that the present coupled-channel $^9$Li~+~$n$~+~$n$ three-body model also explains those observed quantities~\cite{Myo08,Ikeda10}.
The detailed analysis from the experimental side has been summarized in Ref.~\cite{Nortershauser11}.

In the previous works~\cite{Myo07,Myo08}, it has been discussed that the $2p$-$2h$ configurations induced by the tensor and pairing correlations in the $^9$Li core are essential to explain the various physical observables in $^{10}$Li and $^{11}$Li.
It was shown that in $^9$Li, the specific excitations of neutrons into the $p_{1/2}$ orbit are occurred due to the tensor and pairing correlations, so that in $^{10}$Li and $^{11}$Li, the coupling between the excited $p_{1/2}$ neutron in $^9$Li and the last $p_{1/2}$ neutrons give rise to the Pauli-blocking, which plays to lose the total energy of $^{10}$Li and $^{11}$Li with $p$-wave configurations.
As a result, the $s$-wave configurations dynamically gain the energy and are largely mixed, which are sufficient to explain the $s$-wave properties of the two nuclei.
It was found that the tensor correlation gives the stronger effect of the Pauli-blocking than the pairing case~\cite{Myo07,Myo08}.
The large matter and charge radii of $^{11}$Li can be the consequences of the Pauli-blocking effect and becomes the important quantities for the Coulomb breakup reaction, because the charge radius affects the $E1$ transition from the viewpoint of the sum rule value and this transition can be dominant in the Coulomb breakup reaction.
Actually in Ref.~\cite{Myo03} the low-energy $E1$ strength of $^{11}$Li is sensitive to the $s$-wave component of the ground state.
Esbensen {\it et al.} discussed~\cite{Esbensen07} the relation between the Coulomb breakup strength and the charge radius of $^{11}$Li.
They concluded that the simple $^9$Li~+~$n$~+~$n$ model cannot explain the both quantities simultaneously.
This problem is solved in the present coupled-channel $^9$Li~+~$n$~+~$n$ model including the tensor and pairing correlations in the $^9$Li core.

To see the effect of the channel-coupling, we also performed the single channel calculation~\cite{Myo08} using the inert $^9$Li core with the $0p$-$0h$ configuration and adjusting $\hat{V}_\text{core-$n$}$ to fit the experimental $S_{2n}$ value.
The result shows that the $(p_{1/2})^2$ configuration of last two neutrons dominates the $^{11}$Li ground state with the probability as 90.6\%, which preserves the $p$-shell magic number of neutrons.
Instead, the $(s_{1/2})^2$ configuration is mixed very small by only 4.3\%.
Accordingly, the matter and charge radii are obtained as 2.99 fm and 2.34 fm, respectively~\cite{Myo08}, both of which are smaller than the experimental values shown in TABLE~\ref{tab:gspro}.
These results indicate that the halo structure is not so developed and the inert core assumption cannot explain the properties of $^{11}$Li.
For $^{10}$Li, the scattering lengths of $s$-wave states do not show the negative value, which does not support the existence of the virtual $s$-state.
The results of $^{10}$Li and $^{11}$Li in the single channel calculations are significantly different from the results of the coupled $^9$Li~+~$n$~+~$n$ model.
This fact means that the single channel calculation using an inert core is inadequate to understand the structures of $^{10}$Li and $^{11}$Li consistently.

\subsection{Coulomb breakup cross section}
\begin{figure}[b]
\includegraphics[width=8.5cm,clip]{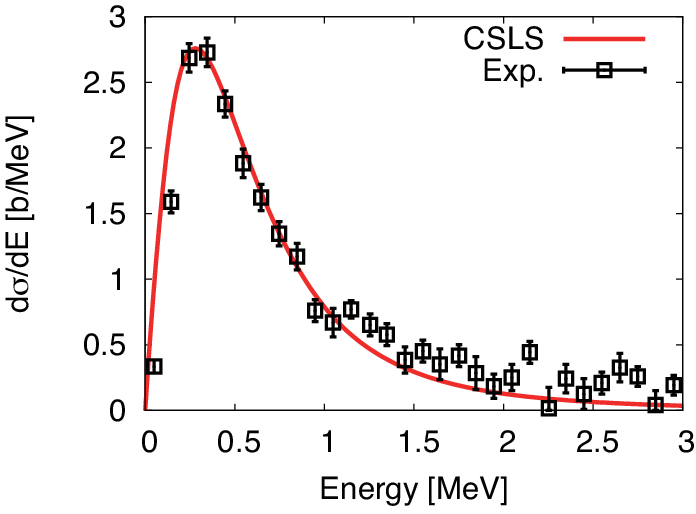}
\caption{\label{fig:cross} (Color online)
Coulomb breakup cross section of $^{11}$Li, measured from the $^9$Li~+~$n$~+~$n$ three-body breakup threshold. The red (solid) line represents the cross sections calculated with CSLS. The experimental data are taken from Re.~\cite{Nakamura07}, shown as the open squares with errorbars.}
\end{figure}

We calculate the Coulomb breakup cross section of $^{11}$Li using Eq.~(\ref{eq:tot_cs}) with the $E1$ strength distribution and taking care of the experimental resolution~\cite{nakamurapri}.
The target is Pb and the incident energy of the $^{11}$Li projectile is 70 MeV/nucleon. 
The cross section measured from the $^9$Li~+~$n$~+~$n$ threshold energy is shown in FIG.~\ref{fig:cross}.
It is found that the results show a good agreement with the experiment~\cite{Nakamura07} for shape and magnitude over the whole energy region.
The distribution shows a low-lying enhancement at around 0.25 MeV and rapidly decreases as the energy increases.
In the CSM calculation, there is no three-body dipole resonance, whose decay width can make a visible structure on the cross section.

\begin{figure}[t]
\includegraphics[width=8.5cm,clip]{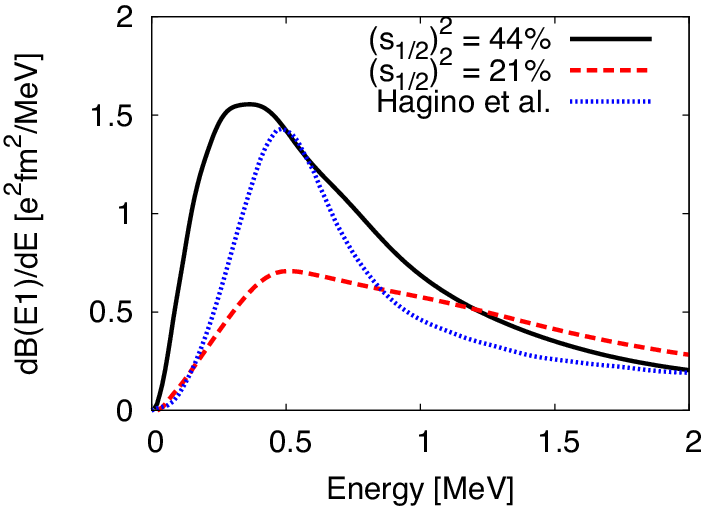}
\caption{\label{fig:dBE1} (Color online)
Comparison between the $E1$ strength distribution with different ($s_{1/2}$)$^2$ components.
The black (solid) line represents the present result used in FIG.~\ref{fig:cross}.
The red (dashed) line is the result using the ground-state wave function with $(s_{1/2})^2=21$ \%.
The blue (dotted) line is the result taken from Ref.~\cite{Hagino09}.
}
\end{figure}

One of the characteristics of $^{11}$Li is quite a large $(s_{1/2})^2$ mixing as 45~\% in the ground state which generates the halo structure. 
It is interesting to see the effect of the large $s$-wave mixing on the Coulomb breakup strength.
We here calculate the $E1$ strength distributions using the different $^{11}$Li wave functions, in which the coupling of the $2p$-$2h$ configuration involved only by the neutron pairing correlation in the $^9$Li core is taken into account.
This restriction of the correlation in $^9$Li leads to the small  $(s_{1/2})^2$ component as 21.0 \% in the ground state~\cite{Myo03}.
The calculated distribution is shown in FIG.~\ref{fig:dBE1} as red (dashed) line.
The distribution shows a relatively small strength at the peak energy, the magnitude of which is about a half of the original one with a large $s$-wave mixing.
The result indicates that a large $s$-wave mixing in the ground state plays a significant role in reproducing the observed low-lying enhancement in the breakup strength.
This point has also been discussed in the previous analysis by changing the $s$-wave properties of $^{11}$Li~\cite{Myo03}.
The explicit effect of the virtual $s$-state of $^{10}$Li in the Coulomb breakup would be discussed using the invariant mass spectra in the next subsection. 
It is suggested that the large $s$-wave mixing enhances the dineutron correlation in the $^{11}$Li ground state~\cite{Myo08}.

It is important to clarify the effect of the correlations in the $^9$Li core on the $E1$ strength distribution.
For this purpose, we compare our coupled-channel calculation including only the pairing correlation, and that of the simple $^9$Li~+~$n$~+~$n$ model assuming an inert $^9$Li core~\cite{Hagino09}, which gives 20.6 \% of the $(s_{1/2})^2$ component and $-5.6$ fm scattering length of the $s$-wave state of $^9$Li-$n$.
The both wave functions contain almost the same amount of the $s$-wave component in the $^{11}$Li ground state.
In two kinds of results, the $E1$ strength distributions commonly have peaks at around 0.5 MeV, however, there exists the large difference of the strength at around the peak energy.
This is due to the fact that about 15 \% of the integrated strength in our calculation escapes to the higher excited $^{11}$Li states having the excited components of the $^9$Li core.

From these comparisons, it is summarized that the large $s$-wave mixing in the initial ground state of $^{11}$Li and the correlations in the $^9$Li core play the essential roles in reproducing the Coulomb breakup cross section, in particular, the position and the magnitude of the low-lying enhancement simultaneously.
Furthermore, the large $s$-wave mixing gives the strong influence on the breakup mechanism.
In our previous work~\cite{Myo03}, we have discussed that the $s$-wave mixing in the ground state enhances the direct breakup process into the $^9$Li~+~$n$~+~$n$ states instead of the sequential one via the $p$-wave resonances in $^{10}$Li.
In fact, when the $s$-wave mixing is about 40\% in the ground state, the three-body direct breakup process exhausts the 66\% of the integrated $E1$ strengths.
This result is much different from the $^6$He case, in which the sequential process via the $^5$He($3/2^-$) resonance dominates the breakup reaction \cite{Myo01}.
This difference between $^{11}$Li and $^6$He can be understood as the effect of two-neutron $s$-wave component on the Coulomb breakup process.
It should be here noticed that the discussion on the breakup mechanism in Ref.~\cite{Myo03} are based on the strength distribution calculated in CSM, in which the $s$-wave virtual states of $^{10}$Li cannot be separated from the continuum states.
Such a calculation cannot distinguish the sequential process via the $s$-wave virtual states from the direct breakup process. 
It is important to derive the invariant mass spectra of the $^9$Li~+~$n$ subsystem in order to estimate appropriately the contribution of the sequential breakup process via the virtual state in the final states, which would be shown in the next subsection.

\begin{figure}[t]
\includegraphics[width=8.5cm,clip]{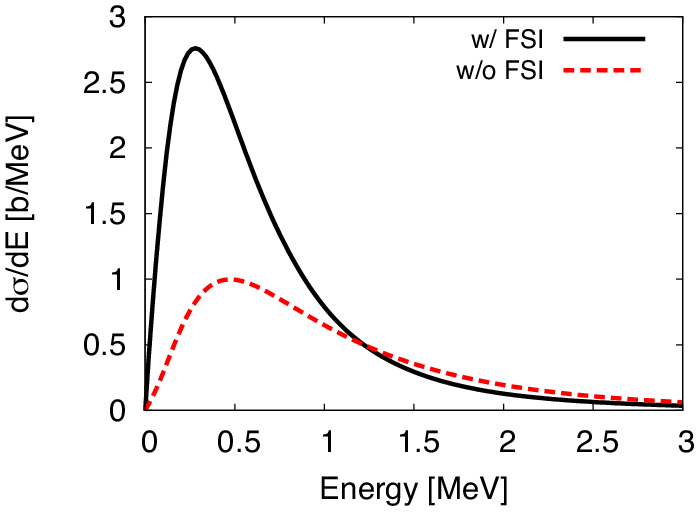}
\caption{\label{fig:cross2} (Color online)
Comparison between the calculated Coulomb breakup cross sections of $^{11}$Li.
The black (solid) line represents the same result in FIG.~\ref{fig:cross}.
The red (dashed) line is the result without FSI.
}
\end{figure}

In addition to the ground-state properties of $^{11}$Li, it is also interesting to see the effect of FSI on the Coulomb breakup cross section.
The FSI is defined in Eq.~(\ref{eq:int}).
In CSLS, all the effects of FSI in the scattering wave functions are included in the second term in Eq.~(\ref{eq:csls}).
We can drop off the second term in the calculation of the cross section to examine the effect of FSI, while we do not change the initial ground-state wave function.
The cross section without FSI is shown in FIG.~\ref{fig:cross2} as red (dashed) line and has a broad peak structure at 0.5 MeV.
It is found that the magnitude is much smaller than the full results including FSI, while there is a very small difference between two results in the higher energy region above 1 MeV.
From this analysis, it is concluded that in addition to the initial state properties of $^{11}$Li, FSI gives a significant effect to create the low-lying enhancement in the Coulomb breakup cross section of $^{11}$Li.
This large effect of FSI has been shown in the previous analyses~\cite{Kikuchi10,Myo08,Hagino09} for the Coulomb breakups of $^{11}$Li and $^6$He.

\subsection{Invariant mass spectra of binary subsystems of $^{11}$Li}
\begin{figure}[b]
\includegraphics[width=8.5cm,clip]{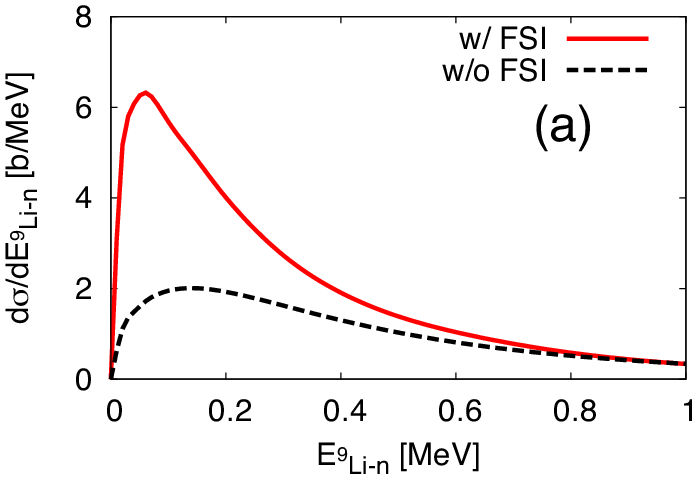}
\includegraphics[width=8.5cm,clip]{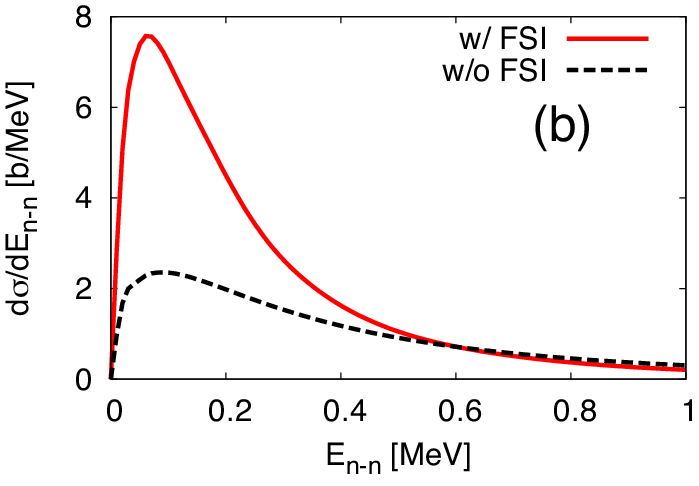}
\caption{\label{fig:invariant}(Color online)
Invariant mass spectra for $^9$Li-$n$ and $n$-$n$ binary subsystems. The panels (a) and (b) represent the results for $^9$Li-$n$ and $n$-$n$ subsystems, respectively. The red (solid) lines show the results with FSI and the black (dashed) ones are those without FSI.}
\end{figure}

To see the effect of FSI in more detail, we calculate the invariant mass spectra using  Eq.~(\ref{eq:inv}) in the Coulomb breakup reaction of $^{11}$Li.
In FIG.~\ref{fig:invariant}, we show the results as functions of the relative energies of $^9$Li-$n$ and $n$-$n$ subsystems as panels (a) and (b), respectively, together with the results calculated without FSI.
It is found that both spectra have sharp peak structures commonly below 0.1 MeV.
From those results, the peaks in the invariant mass spectra are understood to come from FSI.
In FIG.~\ref{fig:invariant} (b), the peak seen in the $n$-$n$ invariant mass spectra is caused obviously by the $n$-$n$ virtual state, because FSI, $\hat{V}_{n\text{-}n}$, produces no resonance.
Such a peak due to the $n$-$n$ virtual state is also seen in the Coulomb breakup reaction of $^6$He~\cite{Kikuchi10}.
\begin{figure}[t]
\includegraphics[width=8.5cm,clip]{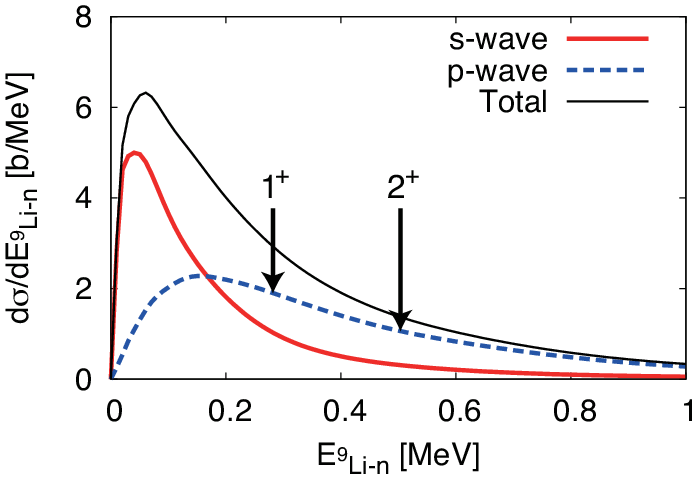}
\caption{\label{fig:inv_decomp}(Color online)
$s$-wave and $p$-wave components of the invariant mass spectra for $^9$Li-$n$. The red (solid) and blue (dashed) lines represent the results for relative $s$- and $p$-wave components, respectively. The black (thin) line is same as the result with FSI in FIG.~\ref{fig:invariant} (a). Two arrows indicate the positions of resonance energies of 1$^+$ and 2$^+$.}
\end{figure}

For the invariant mass spectrum of the $^9$Li-$n$ subsystem, Hagino {\it et al.}~\cite{Hagino09} discussed that both  the $s$-wave virtual state and the $p$-wave resonance of $^{10}$Li contribute to the $E1$ strength distribution using the simple $^9$Li~+~$n$~+~$n$ three-body model without the tensor correlation in the $^9$Li core.
On the other hand, as shown in FIG.~\ref{fig:invariant} (a), our prediction indicates only a single prominent peak below 0.1 MeV.
This peak is considered to come from the $s$-wave virtual state due to $\hat{V}_{\text{core-}n}$.

To see explicitly the partial-wave contributions in the  $^9$Li-$n$ invariant mass spectrum, we calculate the decomposed spectra as shown in FIG.~\ref{fig:inv_decomp}.
The results show that the $s$-wave component has a peak below 0.1 MeV, which comes from the virtual $s$-state of $^{10}$Li.
The $p$-wave component has a broad bump at around 0.15 MeV.
It is found that the bump energy does not correspond to the $p$-wave resonance energies in $^{10}$Li indicated by two arrows in FIG.~\ref{fig:inv_decomp}.
This result means that the $p$-wave contribution in the spectra comes from the non-resonant continuum states of $^9$Li-$n$.
Thus we can conclude that the shape of the invariant mass spectra of $^9$Li-$n$ is mainly determined by the virtual $s$-state in $^{10}$Li, while the non-resonant $p$-wave contribution gives some amount in the spectra, which becomes dominant at higher energies than 0.2 MeV.
This conclusion contradicts the result in Ref.~\cite{Hagino09}, in which the $p$-wave resonance has a sizable contribution to the strength and the virtual $s$-state seems to give a small effect on the strength in comparison with the present result.

The reason why the $p$-wave resonances are not observed in the present $^9$Li-$n$ invariant mass spectra can be understood as follows;
The $p$-wave resonances of $^{10}$Li are located at 0.275 MeV and 0.506 MeV for $1^+$ and $2^+$, respectively, as shown in TABLE~\ref{tab:10Li}.
On the other hand, the breakup cross section has a peak at around 0.25 MeV as shown in FIG.~\ref{fig:cross}.
This peak energy is lower than the energies of the $p$-wave resonances of $^{10}$Li.
The relation of energies implies that the sequential breakup process via the $p$-wave resonances of $^{10}$Li is energetically not favored at around the peak energy of the cross section.
From this energy condition, the $p$-wave resonances give a minor contribution to the Coulomb breakup reaction of $^{11}$Li.
In fact, since the observed breakup cross section shows a peak at 0.25 MeV~\cite{Nakamura07}, it seems to be difficult to observe the $p$-wave resonances in the $^9$Li-$n$ invariant mass spectra if those energies are higher than 0.25 MeV.
On the other hand, the breakup cross section calculated by Hagino {\it et al.}~\cite{Hagino09} has a peak at around 0.5 MeV, locating higher than the $p$-wave resonance energies of $^{10}$Li.
In that case, the sequential breakup via the $p$-wave resonances of $^{10}$Li is favorably allowed, and can make the peak in the strength.

\begin{figure}[t]
\includegraphics[width=8cm,clip]{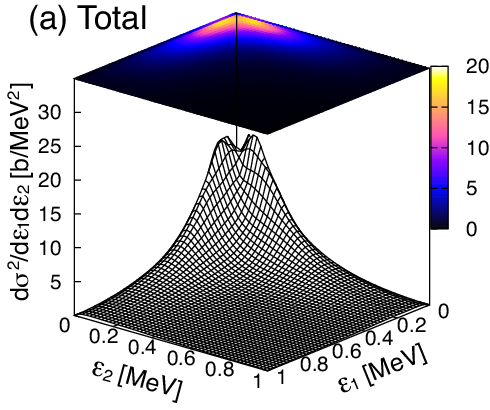}
\vspace{0.3cm}\\
\includegraphics[width=8cm,clip]{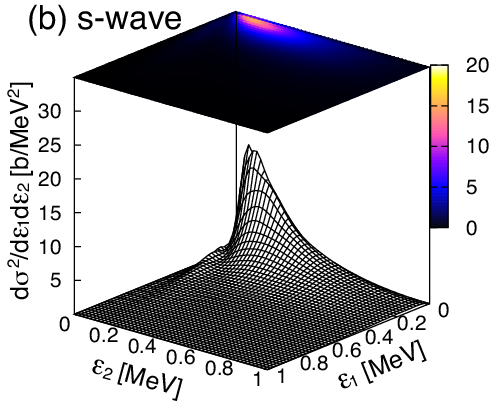}
\vspace{0.3cm}\\
\includegraphics[width=8cm,clip]{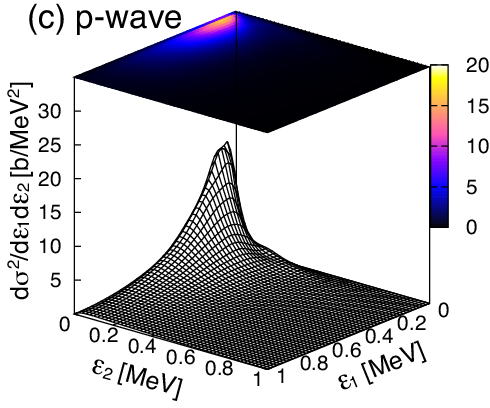}
\caption{\label{fig:2D_wFSI}(Color online)
Two-dimensional energy distributions of the Coulomb breakup cross section of $^{11}$Li, where $\varepsilon_1$ and $\varepsilon_2$ are the relative energy of the $^9$Li-$n$ subsystem and the energy of the relative motion between the center-of-mass of the $^9$Li-$n$ subsystem and the other neutron, respectively.
The panel (a) presents the total distribution, and the panels (b) and (c) present the results for $s$- and $p$-wave components for the relative motion of $^9$Li-$n$, respectively.}
\end{figure}
The contributions of the $s$- and $p$-wave components in the $^9$Li-$n$ invariant mass spectra can be seen more clearly in the two-dimensional energy distributions of the cross section, $d^2\sigma /d\varepsilon_1 d\varepsilon_2$, shown in FIG.~\ref{fig:2D_wFSI}.
Here, $\varepsilon_1$ and $\varepsilon_2$ are the relative energy of the $^9$Li-$n$ subsystem and the energy of the relative motion between the center-of-mass of the $^9$Li-$n$ subsystem and the other neutron, respectively.
The distibutions are calculated by using Eq.~(\ref{eq:2d_cs}).
It is found that the total distribution shown in FIG.~\ref{fig:2D_wFSI} (a) has two peaks at small energies of $\varepsilon_1$ and $\varepsilon_2$, respectively.
The origin of those structures are confirmed by decomposing the distribution into the $s$- and $p$-wave components for the $^9$Li-$n$ states.
In FIG.~\ref{fig:2D_wFSI} (b), the $s$-wave component is shown and the concentration of the strength is found at $\varepsilon_1 = 0.03$ MeV and $\varepsilon_2 = 0.13$ MeV.
This peak near the zero-energy of $\varepsilon_1$ is a consequence of the virtual $s$-state of the $^9$Li-$n$ system.
In FIG.~\ref{fig:2D_wFSI} (c), the $p$-wave component shows a peak at $\varepsilon_1 = 0.12$ MeV and $\varepsilon_2 = 0.03$ MeV and the distribution is relatively wider for $\varepsilon_1$ than the $s$-wave case, which is consistent to the non-resonant breakup of the $p$-wave component in the $^9$Li-$n$ system.
It is also found that the $s$- and $p$-wave components have the similar magnitudes.
Small bumps are obtained at small energies of $\varepsilon_2$ in FIG.~\ref{fig:2D_wFSI} (b) and of $\varepsilon_1$ in FIG.~\ref{fig:2D_wFSI} (c) for the $s$- and $p$-wave components, respectively.
This result suggests a coupling between $s$- and $p$-wave components.

\section{Summary\label{sec:sum}}
We have investigated the Coulomb breakup reaction of two-neutron halo nucleus $^{11}$Li using the coupled-channel $^9$Li~+~$n$~+~$n$ three-body model.
The three-body scattering states of $^{11}$Li are described in terms of the complex-scaled solutions of the Lippmann-Schwinger equation (CSLS).
In the present model of $^{11}$Li, we take into account $2p$-$2h$ configurations of the $^9$Li core, which describes the tensor and pairing correlations in the $^9$Li core on the basis of the tensor-optimized shell model.
As a result, we can reproduce the breaking of the $N=8$ magic number due to the large $s$-wave mixing, which brings the neutron halo structure in $^{11}$Li, and also explains the $s$-wave property of $^{10}$Li simultaneously.
In this paper, we calculated the transition from the $^{11}$Li ground state into the three-body scattering states using CSLS by the Coulomb response. 
In CSLS, the Green's function using the complex-scaled wave functions provides with the three-body scattering states of $^{11}$Li, which include much information of the correlations not only of three-body system but also of the binary subsystems of $^9$Li-$n$ and $n$-$n$.

The calculated Coulomb breakup cross section of $^{11}$Li into the three-body scattering states shows the low-lying enhancement and well reproduces the observed data over whole energy region.
The magnitude of the low-lying enhancement is sensitive to the $s$-wave mixing in the initial $^{11}$Li ground state,
the value of which is obtained as 44\% in the present calculation and reproduces the experimental value.
The excitation of the $^9$Li core is also important to reproduce the total breakup strength, because about 15\% of the strength escapes to the higher energy region as the component of the core excitation in the present coupled-channel approach.
This component reduces the low energy strength corresponding to the experiment. 
These two effects nicely work to determine the distribution of the cross section of $^{11}$Li.

It is confirmed that the low-lying enhancement in the cross section is also affected by the final-state interactions (FSI) in the $^9$Li~+~$n$~+~$n$ three-body system. 
This result is a similar feature to the $^6$He breakup case except for the large $s$-wave mixing in the ground state.
To determine what kinds of FSI dominate the Coulomb breakup reaction of $^{11}$Li, 
we calculate the invariant mass spectra as functions of the energies of the two kinds of the binary subsystems of $^{10}$Li as $^9$Li-$n$, and $n$-$n$.
The calculated spectra show the sharp peak structures below 0.1 MeV from the thresholds in both cases.
It is found that those peaks come from the virtual-state correlations in $^{10}$Li and $n$-$n$ subsystems in the final states, respectively.
On the other hand, it is found that the $p$-wave resonances of $^{10}$Li give a minor effect on the breakup cross section.
This is because the $p$-wave resonances of $^{10}$Li are energetically located higher than the energy of the low-lying enhancement in the breakup cross section, so that the sequential breakup process via the $p$-wave resonances is mostly forbidden in energy.
In relation to this fact, the effect of FSI on the $p$-wave component of the $^9$Li-$n$ invariant mass spectra is suggested to be small.
This feature of the $^{11}$Li breakup is very different from the $^6$He case, which shows a large effect of the $p$-wave component of the $^4$He-$n$ states, and instead, a minor effect of the $s$-wave component.

\begin{acknowledgments}
This work was supported by a Grant-in-Aid for Young Scientists from the Japan Society for the Promotion of Science (No. 24740175).
One of the authors (K. Kat\=o) thanks for support by ```R~\&~D' Platform Formation of Nuclear Reaction Data in Asian Countries (2010-2013)h, Asia-Africa Science Platform Program, Japan Society for the Promotion of Science.
Numerical calculations were performed on a computer system at RCNP, Osaka University.
\end{acknowledgments}

\bibliographystyle{h-physrev}
\bibliography{References}

\end{document}